\documentclass[fleqn,usenatbib]{mnras}

\usepackage[T1]{fontenc}
\usepackage{ae,aecompl}
\usepackage{float}
\usepackage{color}

\usepackage{graphicx}	
\usepackage{amsmath}	
\usepackage{amssymb}	

\usepackage{subcaption}
\usepackage{comment}
\usepackage{xcolor}
\usepackage[b]{esvect}
\usepackage[mathscr]{euscript}

\title[Chaotic secular dynamics]{Instability from high-order resonant chains in wide-separation massive planet systems}

\author[Murphy \& Armitage]{Matthew M. Murphy$^{1,2}$\thanks{E-mail: mmmurphy@email.arizona.edu} and Philip J. Armitage$^{2,3}$
\\
$^{1}$Department of Astronomy and Steward Observatory, University of Arizona, 933 North Cherry Avenue, Tucson, AZ 85721, USA \\
$^{2}$Department of Physics and Astronomy, Stony Brook University, Stony Brook, NY 11790, USA \\
$^{3}$Center for Computational Astrophysics, Flatiron Institute, 162 Fifth Avenue, New York, NY 10010, USA \\
}

\usepackage{newtxtext,newtxmath}
\begin{document}
\label{firstpage}
\pagerange{\pageref{firstpage}--\pageref{lastpage}}
\maketitle

\begin{abstract}
Diversity in the properties of exoplanetary systems arises, in part, from dynamical evolution that occurs after planet formation. We use numerical integrations to explore the relative role of secular and resonant dynamics in the long-term evolution of model planetary systems, made up of three equal mass giant planets on initially eccentric orbits. The range of separations studied is dominated by secular processes, but intersects chains of high-order mean-motion resonances. Over time-scales of $10^8$ orbits, the secular evolution of the simulated systems is predominantly regular. High-order resonant chains, however, can be a significant source of angular momentum deficit (AMD), leading to instability. Using a time-series analysis based on a Hilbert transform, we associate instability with broad islands of chaotic evolution. Previous work has suggested that first-order resonances could modify the AMD of nominally secular systems and facilitate secular chaos. We find that higher-order resonances, when present in chains, can have similar impacts.
\end{abstract}


\section{Introduction}
\label{sec:intro}
Determining the stability of planetary systems is a foundational question in dynamics with multiple applications to Solar System and extrasolar planetary system problems. The two planet case is the only one that is essentially solved \citep{marchal82, gladman93, Petit2018, Hadden2018}. For higher multiplicity, there is a qualitative understanding that chaos and instability result from resonance overlap \citep{lecar01}, and important partial semi-analytic results exist \citep{petit20,yalinewich20, tamayo21}. Numerical integrations, however, are needed to address most questions of astrophysical interest.

The evolution of systems of multiple planets is affected by both secular and resonant processes \citep{murray99}. Under secular evolution, planets exchange angular momentum at fixed energy, leading to changes in orbital eccentricity and mutual inclination while semi-major axes remain constant. The Angular Momentum Deficit (AMD) is conserved in this limit, and if the AMD satisfies ${\rm AMD} < {\rm AMD}_{\rm crit}$, where ${\rm AMD}_{\rm crit}$ is the minimum required for orbit crossing, then the system is ``AMD-stable" \citep{laskar17}. Orbits within an AMD-unstable system, conversely, can potentially evolve until planetary close-encounters occur. Qualitatively different behavior occurs when mean-motion resonances (MMRs) play a significant role. Under resonant dynamics the semi-major axes of the orbits evolve, and non-zero eccentricity can be generated from initially circular and co-planar configurations. Instability arises in such a system due to the overlap of mean-motion \citep{obertas17} and/or three-body resonances \citep{quillen11}. In model systems of equal mass planets that are initially spaced evenly in $\log$(semi-major axis), the average instability time is an approximately exponential function of the planetary spacing \citep{chambers96,smith09,obertas17,lissauer21}, with a dispersion that is set by the chaotic dynamics \citep{hussain20}. 
In general systems, where the masses or separations are non-uniform, good analytic estimates of the instability time have only recently been worked out \citep{petit20}.
The average instability time is a smooth enough function of the system parameters that it can also be estimated reliably using Machine Learning methods \citep{tamayo16,tamayo20,cranmer21}.

The formation and evolution of the Solar System and extrasolar systems is understood to involve a complex mix of secular and resonant dynamics \citep{davies14}, with further complications arising when tides or interactions with gas or planetesimal discs are present \citep{morbidelli18}. In the current Solar System, the dynamics of the planets is controlled by secular effects \citep{laskar90}, and the system is quasi-stable, with a low-probability destabilization of Mercury's orbit remaining as an allowed possibility \citep{laskar94,batygin08,abbot21}. In extrasolar planetary systems, a subset of the currently observed configurations are hypothesized to result from an earlier phase of instability. For systems of giant planets, scattering from near-circular initial conditions -- unstable due to resonant effects -- may explain the observed eccentricity distribution \citep{rasio96,weidenschilling96,ida97,chatterjee08,juric08}. Secular chaos \citep{wu11,teyssandier19}, in multi-planet systems with non-zero initial AMD, is one of several ways to form hot Jupiters via a combination of secular dynamical processes and tidal dissipation \citep{naoz11,petrovich15,dawson18}. At lower masses, the close-in planetary systems that dominate the {\em Kepler}-discovered population \citep{thompson18} may also derive from unstable progenitors, with both resonant and secular dynamics being potentially important \citep{pu15,volk20}. Population analyses suggest that in addition to N-body dynamics, dissipation likely also played a role in forming observed systems \citep{yee21}.

Planetary systems dominated by secular dynamics typically have large separation that exclude the presence of low-order MMRs, and are sometimes studied using numerical schemes that explicitly filter out resonant effects \citep{laskar89}. Physically, however, there could be a class of planetary system where both secular and weaker resonant processes control the stability \citep{wu11,petit17}, and our goal in this paper is to explore the interplay between these processes. We employ a setup that is a variant of the one used by \citet{chambers96} and many subsequent authors: three equal mass planets uniformly spaced in $\log$(semi-major axis). We focus on massive planets, and initialize our systems with moderately large eccentricities. This is evidently a highly idealized system, and we note at the outset that it means that weak resonances in the system always occur in chains between both adjacent pairs of planets rather than between just one pair.

The layout of the paper is as follows. In section \ref{sec:methods} we describe the setup and execution of the integrations done in this work. Our results and interpretations are introduced in section \ref{sec:results} and the implications of these results are discussed in section \ref{sec:discussion}.

\section{Methods}
\label{sec:methods}

In brief, we numerically integrated systems of three massive (Jupiter mass, around a Solar mass star) planets for $10^8$ orbits of the inner planet. The initial conditions were coplanar, with eccentricity set to $e=0.33$ for each planet. In nature, such high eccentricity systems will likely also have significant mutual inclinations, but this coplanar setup allows us to isolate the partitioning of AMD into one variable for ease of analysis. The planets were uniformly spaced in $\log$(semi-major axis), with simulations populating the range of semi-major axis ratios $3.33 \leq \alpha^{-1} \leq 5$, where $\alpha$ is defined as
\begin{equation}
    \alpha \equiv \frac{\mathrm{a}_{\text{inner}}}{\mathrm{a}_{\text{outer}}} \label{eqn:alphadef}
\end{equation}
for a pair of adjacent planets. Integrations were performed using \texttt{REBOUND} \citep{rebound}.

\subsection{Integration Scheme}
\label{subsec:intscheme}

We used \texttt{REBOUND}, an open source N-body integration package developed by \cite{rebound}. \texttt{REBOUND} is equipped with a suite of integrators that differ in their methods and suitability. After comparing and benchmarking \texttt{IAS15} and \texttt{WHFast} on this problem, we adopted the \texttt{WHFast} integrator \citep{wh, reboundwhfast}. To boost performance, we de-toggled the built-in "safe mode" and set the integrator to use an eleventh-order symplectic corrector. 

\subsection{Termination Conditions}
\label{subsec:terminationconds}

Over the course of the integration, pairs of planets may experience collisions, close encounters, and orbit crossings. We terminate runs upon any of these events. For collisions, we set a condition to terminate if the number of planets present in the simulation changes from its initial value. 
We define a close encounter as when the separation between two adjacent planets, $k$ and $k+1$, becomes less than their mutual Hill radius, defined by
\begin{equation}
\mathrm{R}_{\text{Hill, mutual}} \equiv \left( \frac{ \mathrm{m}_k + \mathrm{m}_{k+1} }{3 \mathrm{M}_\star } \right)^{1/3} \frac{ \mathrm{a}_k + \mathrm{a}_{k+1} }{2}. 
\end{equation}
Here, m$_i$ are the planet masses and M$_\star$ is the mass of the star. Lastly, an orbit crossing occurs when the apocentre of planet $k$ becomes exterior to the pericentre of planet $k+1$. Quantitatively, this is when
\begin{equation}
\mathrm{a}_{k+1} \left( 1 - \mathrm{e}_{k+1} \right) < \mathrm{a}_k \left( 1 + \mathrm{e}_k \right).
\end{equation}
Any of these events signify instability for our initially well-separated planets.

\subsection{Initial conditions}
\label{subsec:initconds}

We focus on a model system of three massive planets on initially eccentric, coplanar orbits around a Solar mass star. Each planet is given the same mass of m =  1 M$_\text{J}$ and initial eccentricity of e = 0.33. We set the innermost planet at a$_1 = 1$ AU, then initialise the semi-major axes of the outer planets according to the separation factor $\alpha$. We initially set $\varpi = \Omega = 0$ for each planet, but randomize their initial true anomalies.  This initial alignment in $\varpi$ does not have a significant impact on our results. We found similar results in a suite of test simulations with initially misaligned $\varpi$ which we will discuss in Section~\ref{subsec:initcondcomments}. 

We split the $\alpha$ range into 333 systems, uniformly spaced in log(semi-major axis). It is important to note that defining the separation as in (\ref{eqn:alphadef}) will force any mean-motion resonances to occur in a chain, meaning each pair of adjacent planets will be in the same resonance. These $\alpha$ values are widely spaced enough that we expect secular effects to dominate, and no low-order MMRs to be present, but this range does contain several high-order MMRs. The simulations are purely gravitational, and scale-free apart from the physical collision termination condition. For practical purposes, the results can be re-scaled to different initial orbital radii with an appropriate re-scaling of time units. We ran the integrations for $10^8$ years, equivalent to $10^8$ orbits of the innermost planet, with a timestep of $10^{-3}$ years.  

Under secular exchange of AMD, large peak eccentricities are most readily reached if the inner planets are less massive than the outer planets. By considering equal mass planets, exchange of AMD leads to smaller eccentricity excursions than the unequal mass case that has often been used in simulations of hot Jupiter formation via secular chaos.

\section{Results}
\label{sec:results}

\subsection{Eccentricity Evolution}
\label{subsec:edistributions}
\begin{figure}
    \centering
    \includegraphics[width = \columnwidth]{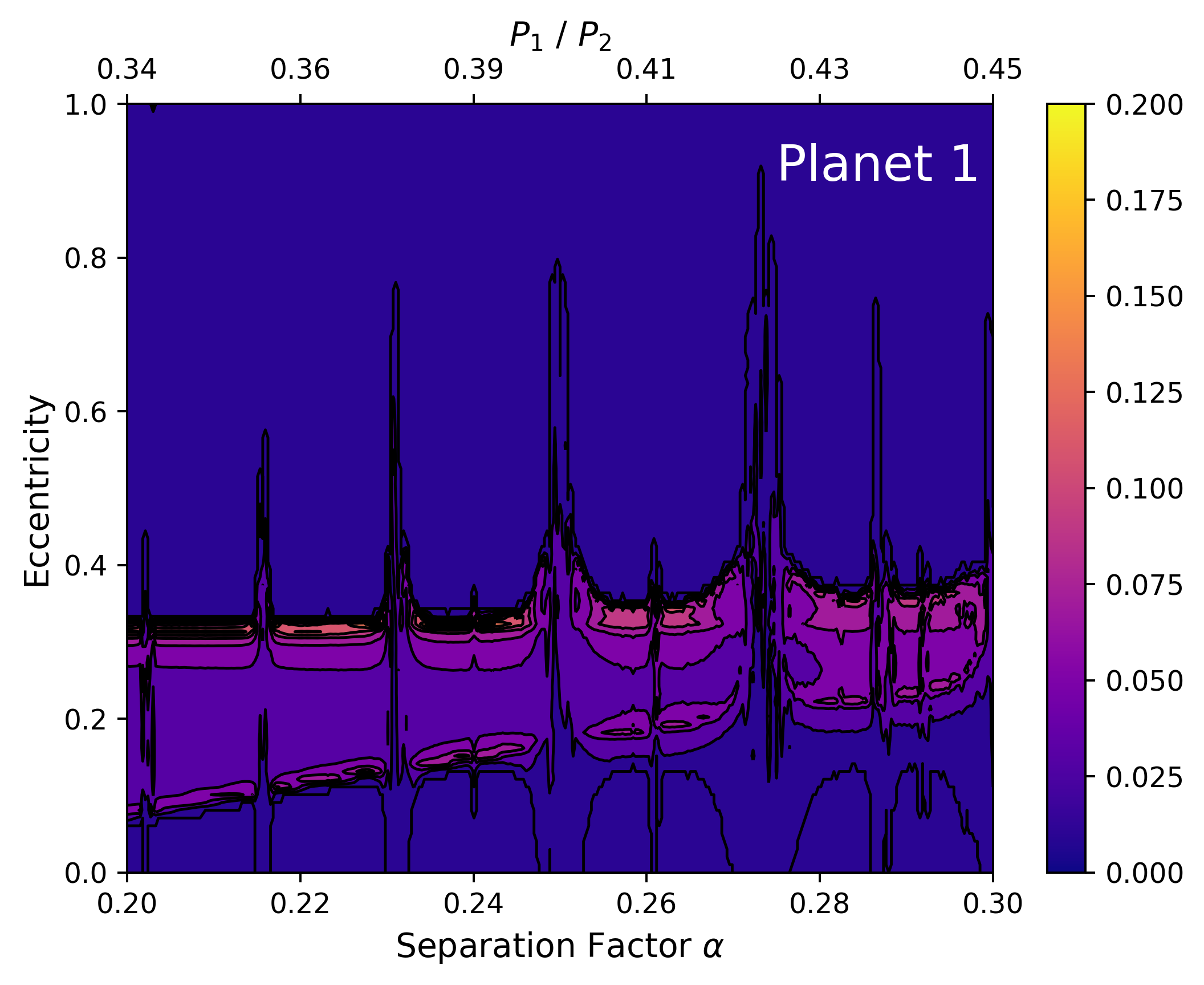}
    \caption{
    Distribution of $\mathscr{P}($e$,\alpha$) for the innermost planet. We also give the ratio of inner and outer planet periods, P$_1$/P$_2$, as a secondary axis. Here, warmer colors indicate a larger value of $\mathscr{P}($e$,\alpha$). The range of eccentricity achieved is widest at lower $\alpha$ and narrows with increasing $\alpha$. There are notable disruptions at $\alpha \approx$ 0.202, 0.215, 0.235, 0.250, 0.261, 0.273, and 0.287. }
    \label{fig:planet1P}
\end{figure}

\begin{figure}
    \centering
    \includegraphics[width = \columnwidth]{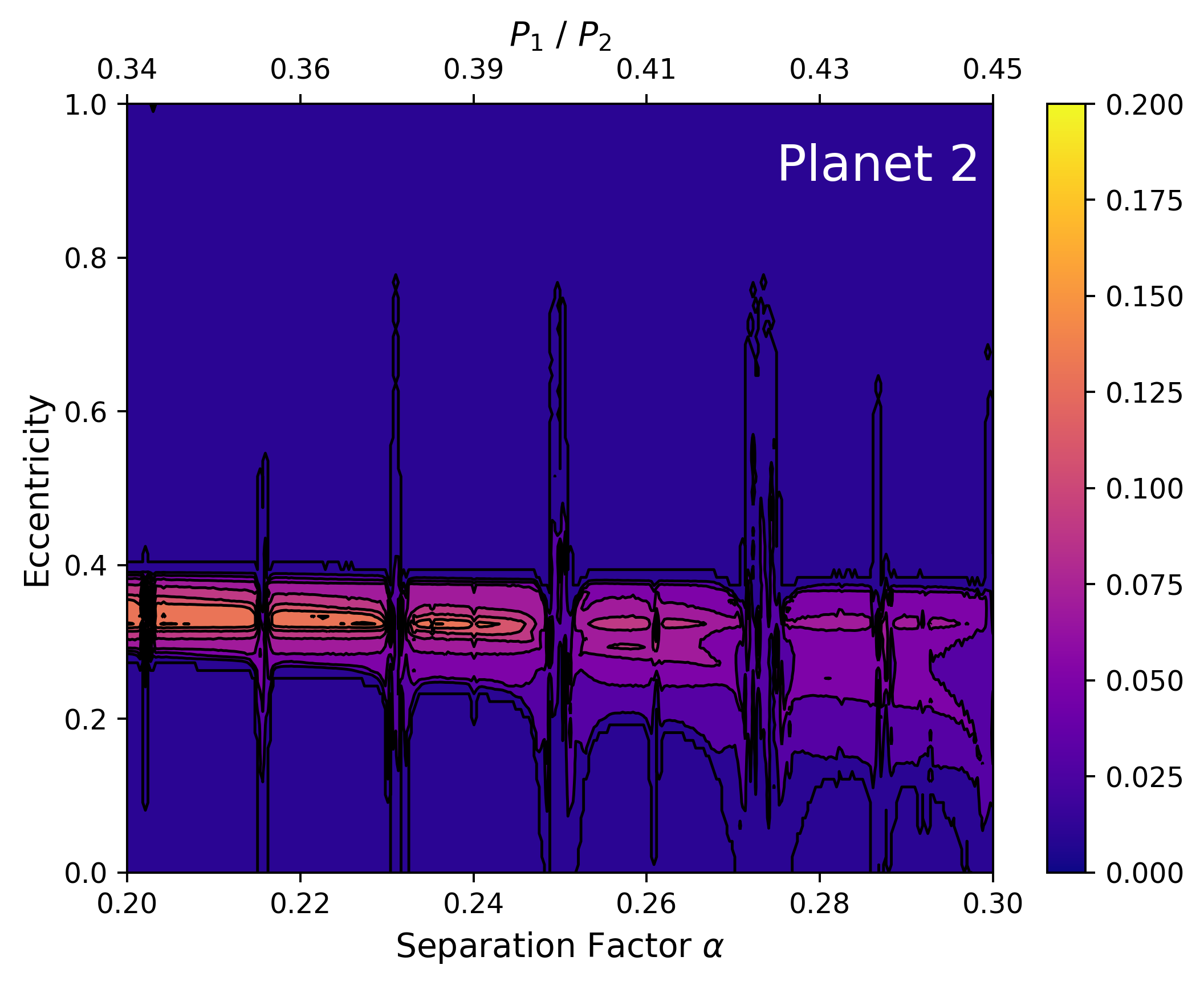}
    \caption{Distribution of $\mathscr{P}($e$,\alpha$) for the middle planet. We also give the ratio of inner and outer planet periods, P$_1$/P$_2$, as a secondary axis. Here, warmer colors indicate a larger value of $\mathscr{P}($e$,\alpha$). The range of eccentricity achieved is most narrow at lower $\alpha$, and widens slightly with increasing $\alpha$. As with that of the innermost planet, there are notable disruptions at $\alpha \approx$ 0.202, 0.215, 0.235, 0.250, 0.261, 0.273, and 0.287.}
    \label{fig:planet2P}
\end{figure}

\begin{figure}
    \centering
    \includegraphics[width = \columnwidth]{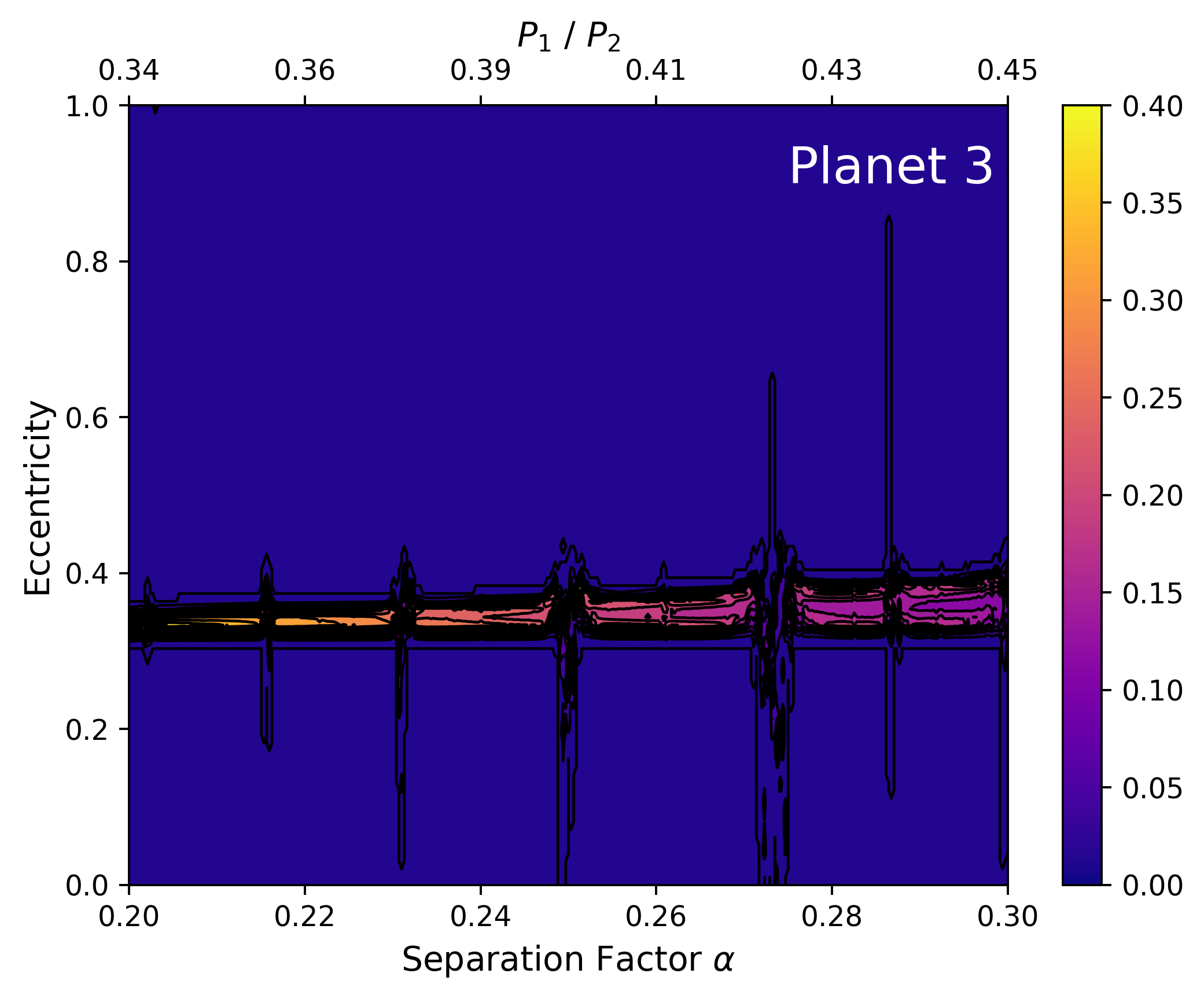}
    \caption{Distribution of $\mathscr{P}($e$,\alpha$) for the outermost planet. We also give the ratio of inner and outer planet periods, P$_1$/P$_2$, as a secondary axis. Here, warmer colors indicate a larger value of $\mathscr{P}$(e$,\alpha$). The range of eccentricity achieved is very narrow at low $\alpha$ which widens very slightly with increasing $\alpha$. As with the other two planets, there are notable disruptions at $\alpha \approx$ 0.202, 0.215, 0.235, 0.250, 0.261, 0.273, and 0.287.}
    \label{fig:planet3P}
\end{figure}

We simulated a range of systems that varied in separation, targeting separation ranges that should largely exhibit secular dynamical evolution. To visualize the ensemble behavior, we rely on a probability distribution-like metric that measures how often each system at each separation was at a particular eccentricity during its integration. We consider 100 eccentricity bins, uniformly spaced across the full range of bound eccentricities $0 \leq$ e $\leq 1$. Then, using the eccentricity time-series for each planet in each system, we evaluate the probability that the planet's eccentricity at any time falls within each bin, then normalize to the total number of data points in the time-series. We denote this distribution as $\mathscr{P}($e$, \alpha)$. Quantitatively, for each system (i.e. each $\alpha$), we compute
\begin{equation}
    \mathscr{P}(e_{\text{bin}}) = \frac{\text{number of counts that fall in bin}}{\text{total number of counts}}. \label{eqn:P(e)bindef}
\end{equation}    
We display these distributions in a two-dimensional representation, with the separation variable $\alpha$ as the other axis, in Figures \ref{fig:planet1P}, \ref{fig:planet2P}, and \ref{fig:planet3P}. We also show the corresponding period ratio between the inner and outer planet in each pair, P$_1$ / P$_2$, as a secondary axis. These plots are color-coded so that brighter and warmer colors represent higher values of $\mathscr{P}($e$,\alpha)$. In systems where the integration terminates due to collision, close approach, or orbit crossing, we evaluate $\mathscr{P}($e$,\alpha)$ from the truncated time series data.

Deterministic secular evolution will lead to periodic, bounded oscillations in eccentricity, such that $\mathscr{P}($e$,\alpha)$ for each planet is a smooth function of $\alpha$. Secular chaos, or departures from secular evolution due to the effects of MMRs, shows up as local (in $\alpha$) deviations from the smooth trend.

    
\subsubsection{Planet Eccentricity Distributions}
\label{subsubsec:differences}

We first examine the $\mathscr{P}($e$,\alpha)$ distribution for the innermost planet, shown in Figure~\ref{fig:planet1P}. The smooth part of the distribution is wider near lower $\alpha$, then narrows as the separation factor increases. This width represents the maximum and minimum eccentricities achieved by these systems. At $\alpha = 0.2$, the distribution ranges between about $0.07 \leq$ e $\leq 0.34$, which narrows to a range $0.25 \leq$ e $\leq 0.4$ at $\alpha = 0.3$. The innermost planet explores a wider range of eccentricity when the system is more widely separated than when it is more compact. In particular, this planet's minimum eccentricity reaches a lower value when the system is more widely separated, while the maximum eccentricity is less affected.  

The innermost planet spends more time at the extremes of its eccentricity distribution. This is evident as the brighter regions and countours along each boundary of the distribution shown in Figure~\ref{fig:planet1P}. The value of $\mathscr{P}($e$,\alpha$) is higher within these contours than within the region between them. This same behavior is also seen in the eccentricity distributions of the Solar System's giant planets, and is a natural realization of quasiperiodic evolution \citep[see e.g.][]{laskar08, mogavero17}. 

The contour near the system's initial eccentricity is wider and has higher $\mathscr{P}($e$,\alpha$) than the other. These converge toward one another as the separation factor increases. Consistent with how the maximum and minimum eccentricities behave, the contour at higher e remains near-stationary in e while the contour at lower e moves significantly toward increasing e. 

The corresponding plots for the middle and outermost planets are shown in Figure~\ref{fig:planet2P} and Figure~\ref{fig:planet3P}. Focusing again on the smooth part of these distributions, the width of the distributions is narrower for these planets, and they are generally centrally peaked. The width of the middle planet's eccentricity distribution broadens toward large $\alpha$ (i.e. more compact configurations). Eccentricity excursions for the outermost planet are only of modest amplitude.

At any given separation, the outermost planet explores a narrower range of eccentricities than the inner two planets. This makes sense in the secular picture. The planets are all equal mass, so the semi-major axis is the leading variable in each planet's angular momentum. As AMD is partitioned between planets, the inner planets are more susceptible to be excited to larger eccentricities, as observed.

\subsubsection{Instabilities in Planet Eccentricity Distributions}
\label{subsubsec:shared}

Departures from smooth secular evolution occur at multiple islands in $\alpha$ that are common across all three planets' eccentricity distributions. At separations of $\alpha \approx$ 0.202, 0.215, 0.235, 0.250, 0.261, 0.273, and 0.287, we see disruptions in the smooth distribution trends where each planet reaches eccentricities beyond the limits of the surrounding trend. These regions occur at the same values of separation and have the same width in $\alpha$ in each distribution, so they are a system-wide phenomenon. The extent of the excursions in eccentricities is not consistent between planets. Whatever causes this disruption affects all planets in the system but not to the same degree. 

\begin{figure*}
    \centering
    \includegraphics[width=\linewidth]{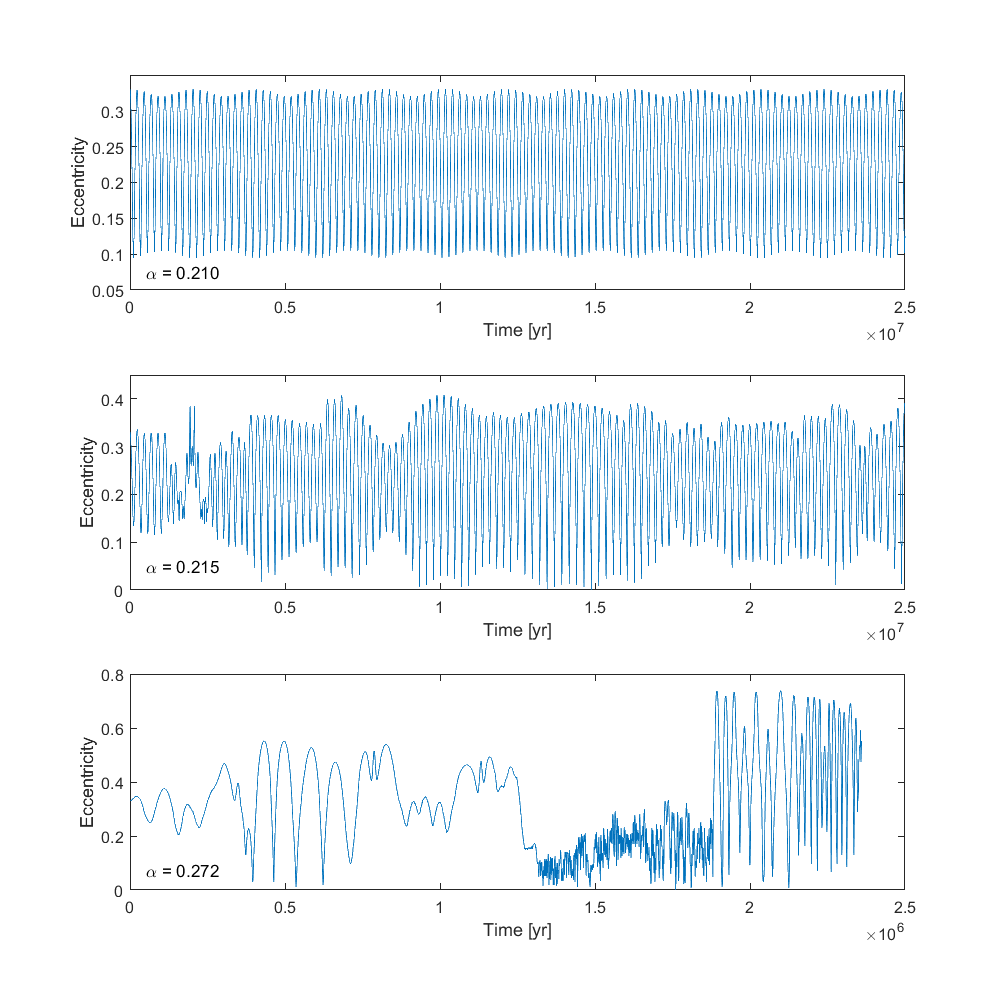}
    \caption{Examples of the e(t) signal in several systems. The top panel shows a 'stable' signal from a system at $\alpha = 0.2102$ which lies outside any disruption region. Notice how the signal exhibits both short-term and long-term periodicity. The middle panel shows a signal from a system at $\alpha = 0.215$ which lies within a disruption region. Notice that the longer-term modulations have been disrupted. The bottom panel shows a signal from a system $\alpha = 0.272$ which lies within a different disruption region. Notice that all periodicity has been disrupted and the system ended with an encounter after 2 Myr}
    \label{fig:e(t)examples}
\end{figure*}

Inspection of individual time series for the simulated systems shows that the evolution can be broadly divided into three classes. Examples are shown in Figure~\ref{fig:e(t)examples}. The top panel of Figure~\ref{fig:e(t)examples} shows a system at $\alpha = 0.2102$, outside any disruption region. This system has stable oscillations in eccentricity, composed of short-period oscillations enveloped by longer-period modulations. The figure is truncated at $10^7$ years but this steady signal continues until the end of integration. The middle panel of Figure \ref{fig:e(t)examples} shows a system at $\alpha = 0.2150$, which is within a disruption region. This system also exhibits steady short-period oscillations but the longer-period modulation is disrupted and has no obvious simple periodicity. The eccentricity amplitude reaches moderately extreme values. This figure has again been truncated at $10^7$ years. Lastly, the bottom panel of Figure~\ref{fig:e(t)examples} shows a system at $\alpha = 0.2722$, within another disruption region. In this system, both the short-term and long-term oscillations have been disrupted and the system ended up having a planetary encounter just before 2.5~Myr. The planet exhibits what appears to be strongly chaotic evolution where the eccentricity attains extreme values as high as near e$=0.8$ and as low as near-zero. 

\subsection{Resonant Interactions}
\label{subsec:resonanceinteractions}

The departures from regular secular evolution shown in Figures~\ref{fig:planet1P} through \ref{fig:planet3P} are presumptively due to a combination of secular chaos and the action of weak mean-motion resonances. Qualitatively, we observe 
two degrees of disruption: one that disrupts only the long-term modulations in eccentricity, and one that disrupts all periodicity of the signal and leads to an encounter. These could be the result of two different processes, or of the same process but to two differing degrees. Not every system within a disruption regions ends with an encounter on the timescale that was tested, but every disrupted system at least exhibits the disruption of long-period modulations, which supports the latter inference. 

\begin{figure}
    \centering
    \includegraphics[width=\columnwidth]{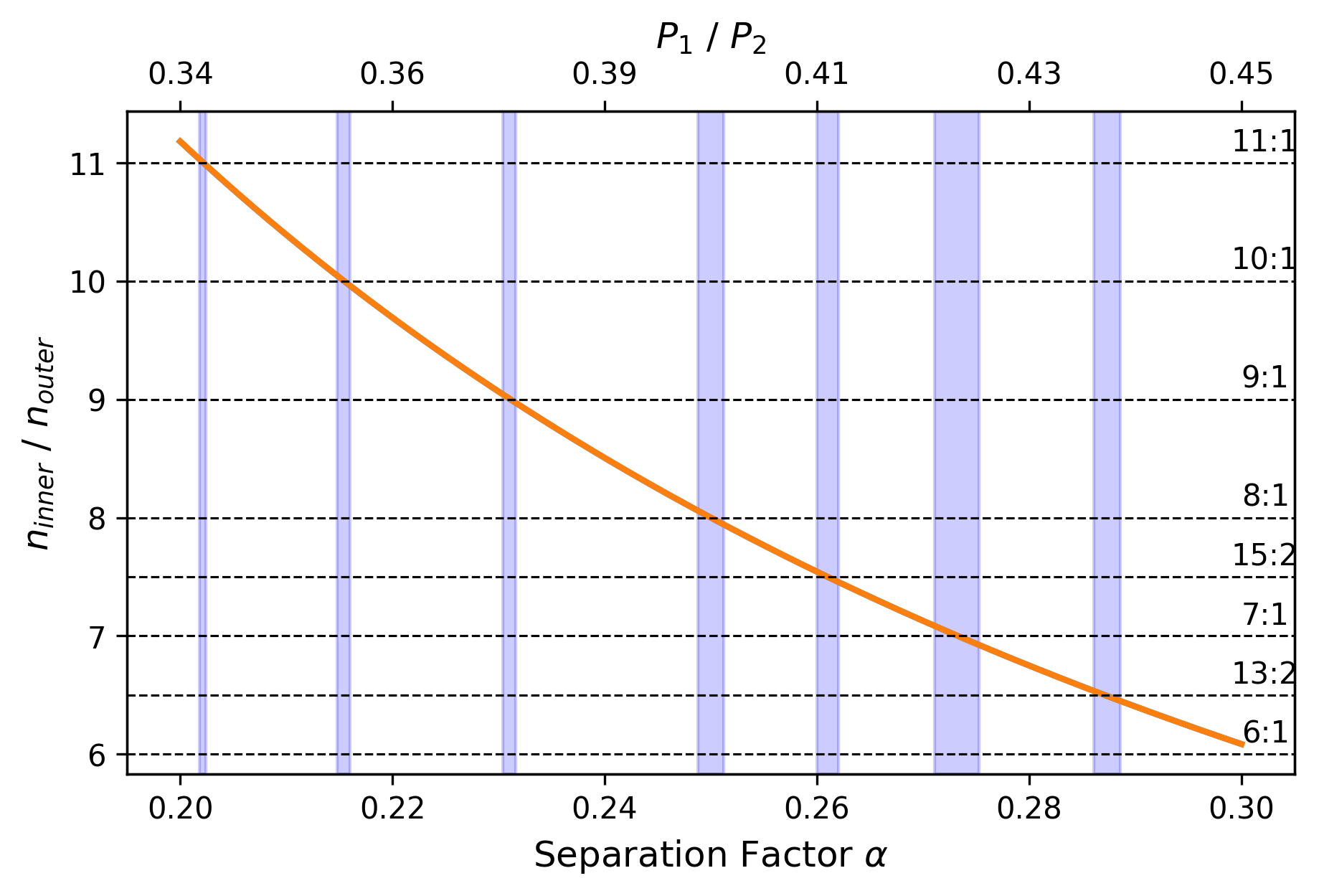}
    \caption{
    Locations of intersections with mean-motion resonances within the range of separations tested. The solid orange curve gives the mean-motion ratio of the planetary pairs as a function of separation. The horizontal dashed lines represent the positions of nearby mean-motion resonances. The horizontal blue shared regions are the locations of the observed disruption regions in Figures \ref{fig:planet1P} - \ref{fig:planet3P}. All but one disruption region intersects exactly with a mean-motion resonance.}
    \label{fig:mmrlocations}
\end{figure}

We computed the ratio of mean-motions for each pair of planets at each separation, which are the same between each pair in each system. The solid curve in Figure~\ref{fig:mmrlocations} shows these mean-motion ratios as a function of separation, with the positions of nearby MMRs overlayed. Also indicated are the ranges of the observed disruption regions, defined based on the distributions shown in Figures~\ref{fig:planet1P}-\ref{fig:planet3P} and checking what range of systems showed disrupted evolutions around each region. 

Nearly all disruption regions align with a mean-motion resonance. From left to right (in Figure~\ref{fig:mmrlocations}), the first region intersects the 11:1 resonance, the second region intersects the 10:1 resonance, the third region intersects the 9:1 resonance, the fourth region intersects the 8:1 resonance, the fifth region intersects the 15:2 resonance, the sixth region intersects the 7:1 resonance, and the seventh region intersects the 13:2 resonance. Each intersected MMR is a chain resonance, meaning that each planet in the system is in resonance with its adjacent neighbor. Evidently, these MMR chains contribute to the observed disruption. These are all high-order resonances and are relatively weak compared to low-order resonances commonly seen in planetary systems. However, the fact that both adjacent pairs of planets share the resonance may significantly strengthen their effect.
    
\begin{figure}
    \centering
    \includegraphics[width=\columnwidth]{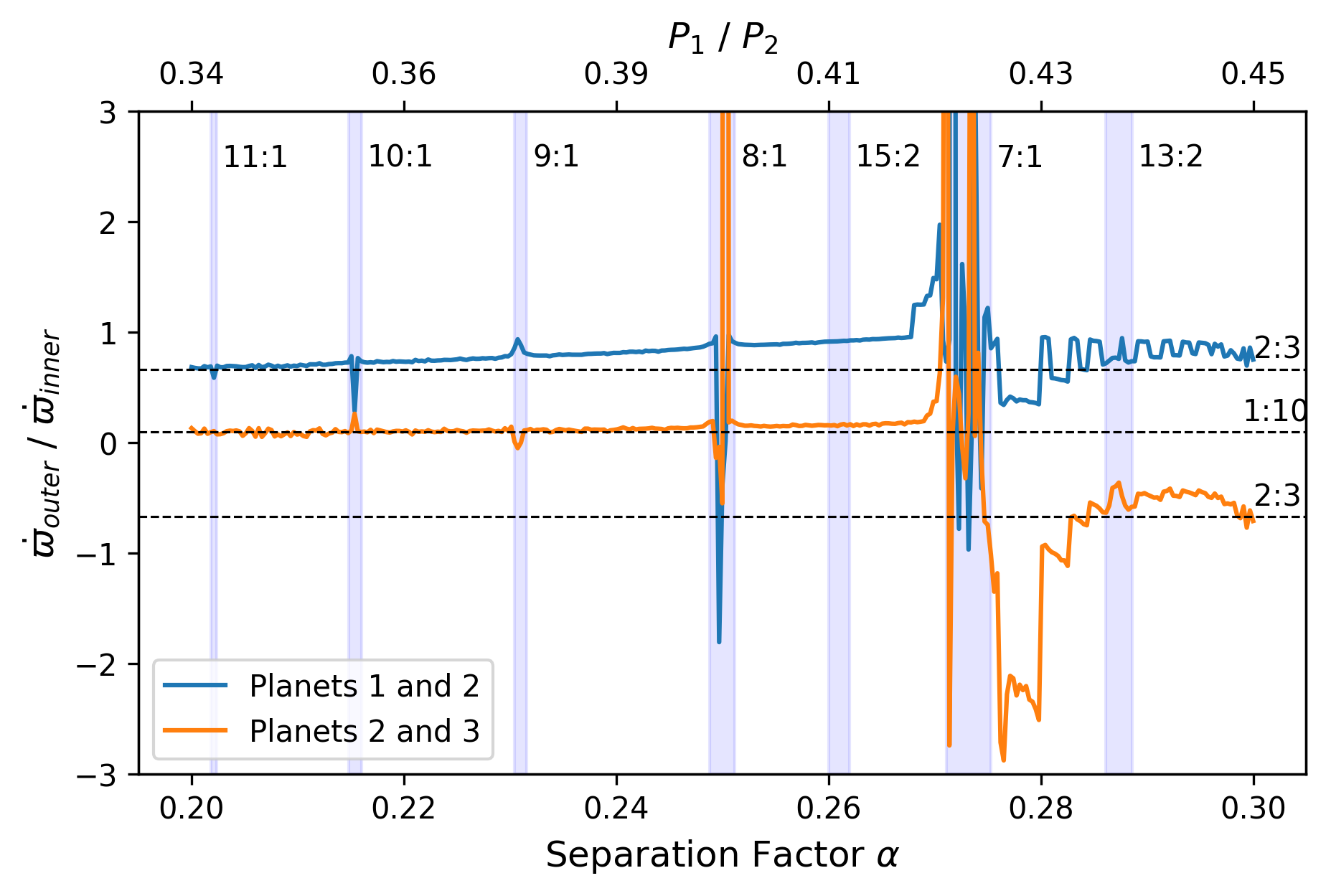}
    \caption{ 
    Locations of secular resonances within the range of separations tested. The solid blue curve represents the ratio of $\varpi$ precession rates for the innermost pair of planets, while the solid orange curve is the same for the outermost pair of planets. The horizontal dashed lines represent the positions of nearby secular resonances. The arrows point to the positions of the observed disruption regions. Notice that the solid curves lie along secular resonances across most of the separation range. At most disruption regions, the curves are deflected off the resonance line.}
    \label{fig:precessions}
\end{figure}

Secular resonances occur when there is a commensurability in the $\varpi$ precession rates of a pair of planets. The $\varpi$ time series for each planet in each system exhibits near-linear circulation from -$\pi$ to $\pi$. We computed the precession rate as the slope of the first linear portion of each system's time series. In Figure~\ref{fig:precessions}, we plot the ratio of these precession rates for each adjacent pair of planets as a function of separation. The positions of nearby secular resonances and the observed disruption regions are indicated. 

Both pairs of planets lie along a secular resonance for nearly the entire range of separations. The innermost pair are in a first order 2:3 resonance and the outermost pair are in a high-order 1:10 resonance. When a disruption region is entered, and correspondingly an MMR is intersected, the pairs appear to be knocked off from their secular resonance. This is first evident at the separation corresponding to the 11:1 MMR region and continues up to the 7:1 MMR region, but does not seem to occur for the 15:2 or 13:2 MMR. The degree to which the planets are knocked from their secular resonance mostly increases with the increasing strength (i.e. lower order) of the intersected MMR, with the 7:1 MMR region exhibiting the strongest effect. After this region, the outermost pair strays from the 1:10 secular resonance and instead continues to higher $\alpha$ along the negative 2:3 resonance. 

Comparing Figures~\ref{fig:planet1P}-\ref{fig:planet3P} and \ref{fig:e(t)examples} shows that systems within the observed disruption regions have partially to fully disturbed eccentricity time series, and those outside disruption regions exhibit stable evolution. Figures \ref{fig:mmrlocations} and \ref{fig:precessions} show that systems within the disruption regions are in the chaotic space surrounding an MMR chain while those outside disruption regions are only in a secular resonance. All together, this suggests that the MMRs cause disrupted eccentricity evolution whereas those systems only in secular resonance are stable. The degree of instability is correlated to the order of the MMR. At all intersections, at least the long-term modulations in e are disrupted and the magnitude of e is driven to relatively extreme values. The higher order MMR chains are unable to fully dominate the stability enforced by the secular resonance, so these are only able to disrupt the longer-period secular modulations. As the MMRs become stronger, they become more dominant. The lower order MMRs generally affect a wider width in $\alpha$, more fully push the planets out of secular resonance, disrupt even short-term periodicity, and sometimes enable planetary encounters.

\begin{figure}
    \centering
    \includegraphics[width=\columnwidth]{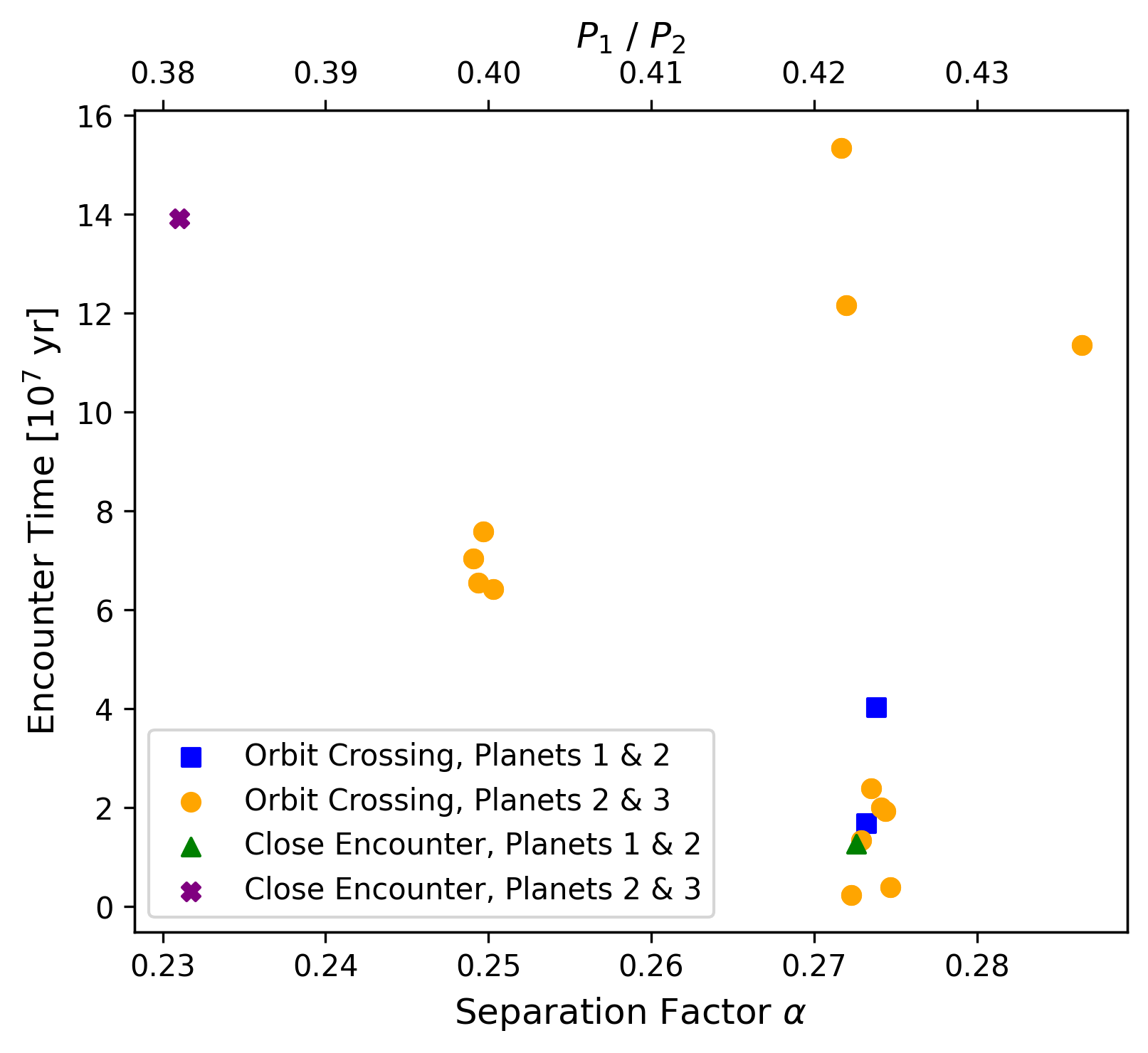}
    \caption{Distribution of encounter times and types as function of the system's separation factor.}
    \label{fig:instability_times}
\end{figure}

A subset of the systems experience encounters over the duration of the simulations. The distribution of encounter times is shown in Figure~\ref{fig:instability_times}. There are two major groupings around $\alpha = 0.25$ and $\alpha = 0.273$ that coincide with the 8:1 and 7:1 MMR chains, which are the two strongest MMRs intersected. 15 of the 17 encounters were an orbit crossing, with 13 of these being between the outermost pair of planets.

\begin{figure}
    \centering
    \includegraphics[width=\columnwidth]{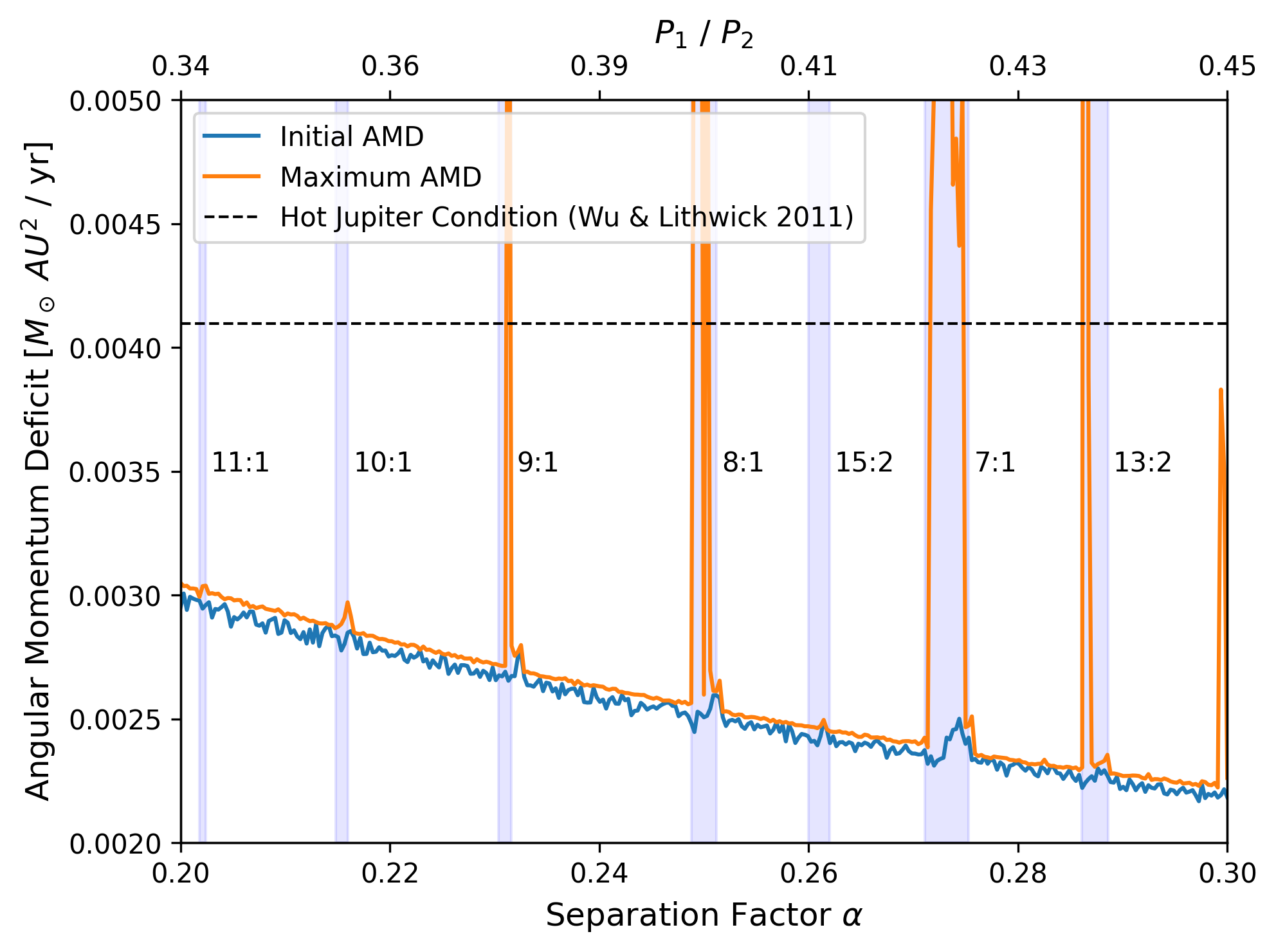}
    \caption{Values of the angular momentum deficit within our systems as a function of separation. The blue curve represents the initial AMD in each system and the orange curve represents the maximum AMD achieved by that system during the integration. The dotted line represents the Hot Jupiter stability condition given in equation (\ref{eqn:HJamdcondition}). The vertical shaded regions represent the positions of the observed MMR regions.}
    \label{fig:amd}
\end{figure}

The angular momentum deficit can be used as an indicator of stability \citep[e.g.][]{laskar17, petit17, wu11, tamayo21}. System AMD is conserved in the secular limit, so the initial value of AMD can forecast the future stability of a purely secular system. Secular chaos can drive massive planets to become hot Jupiters through eccentricity excitation followed by orbital decay and then eventual tidal capture by the star. This defines one limit of AMD stability: the amount of AMD that, if transferred in full to the innermost planet, would lead it to become a hot Jupiter. For a coplanar system, assuming that the orbit becomes tidally circularized by the star at $\mathrm{a}_1 \left( 1 - \mathrm{e}_1 \right) \leq 0.05~AU$, this condition is \citep{wu11}
\begin{equation}
\mathrm{AMD} \geq \Lambda_1 \left( 1 - \sqrt{ \frac{ 0.1~AU }{\mathrm{a}_1} } \right), \label{eqn:HJamdcondition}
\end{equation}
where $\Lambda_k = \frac{\mathrm{m}_k \mathrm{M}_\star}{\mathrm{m}_k + \mathrm{M}_\star} \sqrt{ G \left( \mathrm{M}_\star + \mathrm{m}_k \right) \mathrm{a}_k }$ is the circular angular momentum of the planet. 
The presence of non-secular processes like nearby MMRs can be a source of additional AMD \citep{wu11}, violating AMD conservation. We find that the MMR chains present in our simulated systems do lead to the violation of AMD conservation. To show this, we plot the initial AMD in each system as well as the maximum value of AMD ever achieved by that system in Figure~\ref{fig:amd}. The horizontal dotted line is the hot Jupiter condition of equation (\ref{eqn:HJamdcondition}). The initial AMD slowly decreases with decreasing separation, which is to be expected. 
The initial AMD of every system is below the condition for the onset of hot Jupiter formation, so these systems are all initially AMD-stable. In most systems, the AMD is conserved throughout the integration as the maximum AMD achieved tracks the initial AMD. However, there are several spikes in the maximum AMD curve that align with the intersected mean-motion resonance chains, indicated as the shaded regions in Figure \ref{fig:amd}. Within the intersections of the 9:1, 8:1, 7:1, and 13:2 MMR chains, the maximum AMD surpasses the condition for the onset of hot Jupiter formation. This shows that chains of high order mean-motion resonances can be a significant source of AMD in widely separated systems and lead to instability in an initially AMD-stable system, albeit only for a small region of parameter space.

\subsection{Time Series Analysis}





The defining feature of chaos is a loss of direct determinism, due to the extreme sensitivity of the final state of the system on its initial conditions. To identify the presence of chaos in the simulated systems, we employ an analysis that is based upon a Hilbert transform of the time series data \citep{liu10}. The Hilbert transform takes a real time series, $q(t)$, and computes a corresponding imaginary time series, $\Tilde{q}(t)$, through the transformation 
\begin{equation}
    \tilde{q} (t) =  P.V. \left[ \frac{1}{\pi} \int_{- \infty }^{\infty } \frac{ q (t^{'} ) }{ t - t^{'} } d t^{'} \right] ,
\end{equation}
where P.V. represents the Cauchy principal value. Then, the real and imaginary series form a so-called \textit{analytic} signal, $\psi(t)$, which can be written as a rotation in complex space as
\begin{equation} 
\psi (t) = q (t) + i \Tilde{q} (t) = A(t) e^{i \phi (t) } .
\end{equation} 
The amplitude is
\begin{equation}
    A(t) = \sqrt{ {q(t)}^2 + {\Tilde{q}(t)}^2 },
\end{equation}
and the phase variable is \citep{liu10}
\begin{equation}
    \phi (t) = \arctan \left( \frac{ \Tilde{q} (t) } { q (t) } \right) .
\end{equation}
Chaos can show up as either irregular modulations in the amplitude or phase of the signal. By transforming into Hilbert space and determining the amplitude and phase variable of the analytic signal, we separate and can independently characterize the effects of amplitude and phase modulation. We expect that the components for stable and chaotic systems will exhibit quantitatively different behavior.

\begin{figure}
    \centering
    \includegraphics[width=\columnwidth]{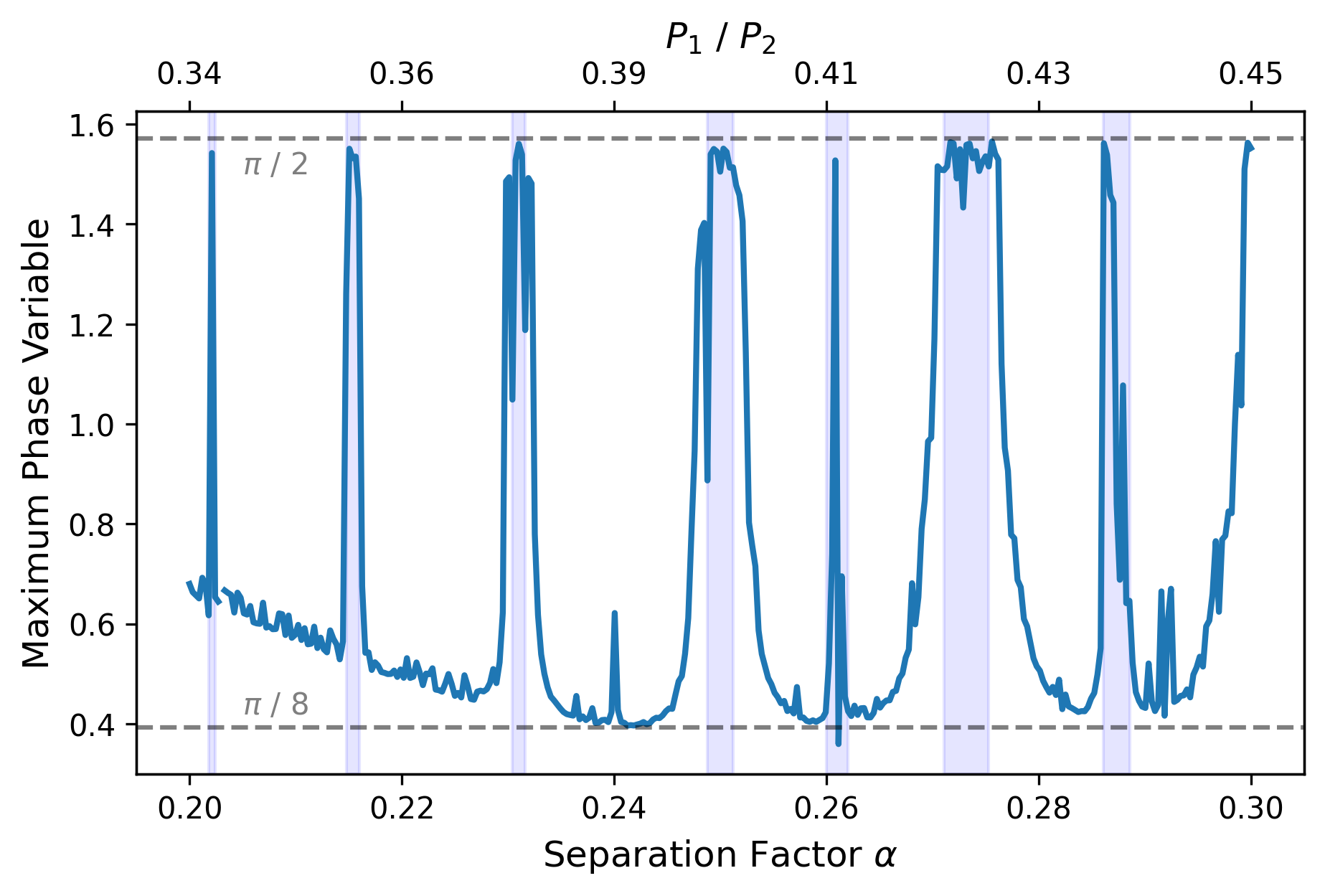}
    \caption{The maximum value achieved by the Hilbert transform phase variable in each system. In systems which intersect a mean motion resonance chain, the phase variable reaches upwards of $\pi$/2, whereas systems outside of mean motion resonance remain bounded at lower values.}
    \label{fig:maxphases}
\end{figure} 

We computed the Hilbert transformation for the inner planet's eccentricity time series for each system. The amplitude series resemble that of the eccentricity series, so their distributions reflect Figures~\ref{fig:planet1P}-\ref{fig:planet3P} and do not reveal new information. The phase variables, however, show significant differences. Each phase variable series librates in time. We plot the maximum phase achieved by each series, representing the limits of that libration, in Figure~\ref{fig:maxphases}. We see markedly different behavior between the stable systems and the systems which intersect an MMR chain. The systems which intersect the chaotic region around an MMR approach a maximum limit phase of $\pi / 2$. The stable achieve a lower limit, around $0.7$ for low $\alpha$ which decreases to a limit of $\approx \pi / 8$ as $\alpha$ increases. 

The eccentricity phase variable of the stable systems is unwrapped, meaning the eccentricity evolution behaves as a multi-valued function. Slightly different input conditions can, in general, evolve to the same end states. 
In contrast, we find that the MMR intersecting systems have a wrapped phase variable with a principal value of $\pi / 2$. This is an indicator of chaos, as each possible initial condition of such a system will have a near-unique end state.

\section{Summary \& Discussion}
\label{sec:discussion}

In this work, we set out to investigate the interplay between secular and resonant interactions in model systems of three eccentric, equal mass (1~M$_{\rm J}$, for a Solar mass host), widely separated planets. We simulated 333 coplanar systems in which the planets were initially separated by factors of 3.33-5 in semi-major axes. A summary of our findings is as follows:
\begin{itemize}
    \item
    Systems that were not in mean-motion resonance remained stable for the duration of the integrations ($10^8$ orbits of the innermost planet). Evolution of these systems was consistent with deterministic secular dynamics.
    \item 
   Systems within the chaotic region surrounding high-order mean-motion resonant chains became unstable, evidenced primarily by a loss of periodic evolution. The stronger mean-motion resonances also led to planetary encounters, including orbit crossings and close encounters. 
    \item
    The high-order mean-motion resonant chains provided a significant source of angular momentum deficit, that was sufficient in many cases to drive initially AMD-stable systems to become unstable.   
\end{itemize}

\subsection{Comments on the Initial Conditions}
\label{subsec:initcondcomments}

We chose to initialise our systems with aligned periapsis longitudes, such that $\varpi_{init}$ = 0 for all planets. This choice did not have a significant impact on the results of our simulations. To test the effect of this choice, we ran additional simulations in which adjacent planets had anti-aligned periapsis longitudes. The initially anti-aligned systems evolved nearly the same as the initially aligned systems. The only major difference was that, in some cases, the limits of a planet's eccentricity time-series where shifted -- e.g. in the aligned case, the eccentricity oscillated between the initial value and a minimum with some amplitude and, in the anti-aligned case, the eccentricity oscillated between the initial value and a maximum instead with the same amplitude. Also, we specifically tested systems at the boundaries of the chaotic regions surrounding the MMRs and found that switching to the anti-aligned case did not extend these regions in the $\alpha$ direction.

The even spacing of the planets in $\log$(semi-major axis), means that we cannot generalize our results to arbitrary massive planets evolving in the nominally secular regime. Instability driven solely by secular processes is well studied, particularly in the context of hot Jupiter formation \citep[see e.g.][]{wu11,lithwick13, laskar17,teyssandier19}. The majority of these studies have explicitly avoided the presence of MMRs, while some have admitted first-order MMRs. \cite{wu11} showed that a first-order MMR aided AMD diffusion between planets and proposed that such resonances could be a significant source of AMD. \cite{petit17} expanded the framework of AMD-stability to account for the effects of first-order MMRs and found that, when applied to observed exoplanetary systems, a small percentage of otherwise AMD-stable systems were actually unstable due to the presence of mean-motion resonances. Our work supports the hypothesis that higher-order resonant interactions may also contribute to instability. 

\subsection{Discussion of Results and Implications}
\label{subsec:implications}
We have found that secular and resonant processes are jointly important for determining the stability of the class of systems that we simulated. Among observed exoplanets, there are examples of systems whose dynamical stability results from the presence of both secular and mean-motion resonances. A combination of secular resonance and a first-order 2:1 MMR contribute to stability in the HD 160691 system \citep{kiselva04}, the GJ 876 system \citep{lee02}, and the HD 82943 system \citep{Hadj03}. Similarly, secular resonance in addition to a nearby high-order 11:2 MMR contribute to the stability of the HD 12661 system \citep{Gozd03}. On the other hand, there are also examples of exoplanetary systems where both secular and mean-motion resonances contribute to instability and chaotic orbits. This has been particularly observed in planets in binary star systems \citep{satyal14,sutherland19}. Considerably less work has been done on systems in which stabilizing secular interactions are overcome by MMRs. This type of interaction has been shown for small bodies within the Solar System, such as the Quadrantid meteors \citep{sambarov20}. Our work has shown that such an interaction can occur between planets in a massive, widely separated, initially eccentric system. 

Determining the space of initial conditions that would evolve under dynamical processes to form the observed population of planetary systems remains an open theoretical challenge. Within the framework of the Nice model \citep{tsiganis05}, statistical constraints have been placed on the initial configuration of the outer Solar System \citep{nesvorny12,clement21}, and current theoretical tools \citep{tamayo20,cranmer21} may be adequate to enable similar studies of the population of compact low-mass planetary systems, where instability is driven primarily by resonant dynamics. More work, however, is needed to quantify the interplay of secular and resonant effects in more general classes of planetary system, and from that to infer constraints on planet formation models from observed exoplanet populations.

\section*{Data availability}

The original data underlying this article are available upon reasonable request to the authors.

\section*{Acknowledgements}

The authors would like to thank our anonymous journal reviewer and scientific editor for their positive feedback and useful suggestions which helped improve this work. The authors would also like to thank Stony Brook Research Computing and Cyberinfrastructure, and the Institute for Advanced Computational Science at Stony Brook University for access to the high-performance SeaWulf computing system, which was made possible by a \$1.4M National Science Foundation grant (\#1531492). 

This work was partially supported with funding from the Stony Brook University Undergraduate Research \& Creative Activities program. Simulations in this paper made use of the REBOUND N-body code \citep{rebound}. The simulations were integrated using WHFast, a symplectic Wisdom-Holman integrator \citep{reboundwhfast,wh}.

\bibliographystyle{mnras}
\bibliography{ref}

\end{document}